\documentclass[10pt,letterpaper,twocolumn,prl,showpacs]{revtex4-1}

\usepackage[english]{babel}

\usepackage{amsmath,amssymb}
\usepackage[dvips]{graphicx,color}

\begin{document}

\title{Controlled merging and annihilation of localized dissipative structures in an AC-driven damped nonlinear Schr\"odinger system}

\author{Jae K. Jang$^1$}
\author{Miro Erkintalo$^1$}
\author{Kathy Luo$^1$}
\author{Gian-Luca Oppo$^2$}
\author{St\'ephane Coen$^1$}
\author{Stuart G. Murdoch$^1$}

\affiliation{$^1$Dodd-Walls Centre and Department of Physics, The University of Auckland, Private Bag 92019, Auckland 1142, New Zealand,}
\affiliation{$^2$SUPA and Department of Physics, University of Strathclyde, Glasgow G4 0NG, Scotland}

\begin{abstract}
  \noindent We report studies of controlled interactions of localized dissipative structures in a system described
  by the AC-driven damped nonlinear Schr\"odinger equation. Extensive numerical simulations reveal a diversity of
  interaction scenarios that are governed by the properties of the system driver. In our experiments, performed
  with a nonlinear optical Kerr resonator, the phase profile of the driver is used to induce interactions on
  demand. We observe both merging and annihilation of localized structures, i.e., interactions governed by the
  dissipative, out-of-equilibrium nature of the system.
\end{abstract}

\maketitle

\noindent Localized structures coexisting with a homogeneous background are ubiquitous phenomena in extended
dissipative systems driven far from equilibrium. These structures consist of solitary excitations that manifest
themselves as electrical pulses in nerves \cite{hodgkin_quantitative_1952}, concentration spots in chemical reactions
\cite{pearson_complex_1993, *lee_pattern_1993}, oscillons in water waves \cite{wu_observation_1984,
clerc_soliton_2009} and in granular matter \cite{umbanhowar_localized_1996, *lioubashevski_dissipative_1996},
filaments in gas discharges \cite{astrov_formation_1997, *astrov_plasma_2001}, patches and fairy circles in
vegetation \cite{lejeune_localized_2002, *fernandez-oto_strong_2014}, or feedback and cavity solitons in nonlinear
optics \cite{taranenko_spatial_1997, *schapers_interaction_2000, barland_cavity_2002, *ackemann_chapter_2009,
leo_temporal_2010, *jang_ultraweak_2013, *herr_temporal_2014}. More generally, they are referred to as localized
dissipative structures (LDSs) or dissipative solitons \cite{akhmediev_dissipative_2008, purwins_dissipative_2010}.

Like the conventional solitons of conservative integrable systems, LDSs can interact and collide with each other,
sometimes with particle-like characteristics. But while conventional solitons always emerge unscathed from
collisions \cite{zakharov_interaction_1973}, LDSs can form bound states, merge into one, or even
annihilate~\cite{purwins_dissipative_2010}. These complex interactions arise from the non-integrability of nonlinear
dissipative systems, and their study is of particular interest to better understand systems outside thermal
equilibrium. Merging and annihilation of solitons have been extensively studied experimentally in non-integrable
\textit{conservative} systems, mostly with optical waves \cite{shih_three-dimensional_1997,
*rotschild_long-range_2006, tikhonenko_three_1996, *shih_incoherent_1996, *krolikowski_fusion_1997,
*krolikowski_annihilation_1998}, but also, more recently, with matter waves \cite{nguyen_collisions_2014}. In
contrast, although several authors have reported complex behaviors of ensembles of LDSs in various settings,
experimental observations have been uncontrolled and mostly qualitative (see, e.g., \cite{umbanhowar_localized_1996,
*lioubashevski_dissipative_1996, purwins_dissipative_2010}). It is only in gas discharges
\cite{bodeker_measuring_2004} and in vertically-driven fluids \cite{clerc_soliton_2009} that quantitative
measurements of the interaction laws have been obtained, with \cite{clerc_soliton_2009} also resolving the merging
dynamics. These two latter examples are realizations of, respectively, a reaction-diffusion system and a
parametrically-driven damped nonlinear Schr\"odinger equation (NLSE) near the 2\,:\,1 resonance.

Here we report on a detailed numerical and experimental study of controlled merging and annihilation dynamics of LDSs
in a system described by an AC-driven damped NLSE near the 1\,:\,1 resonance. Experiments are performed in a
nonlinear optical Kerr resonator, in which we can excite LDSs at selected and precise positions, and systematically
induce their interactions. The interactions are triggered by manipulating the phase profile of the driver; the
outcome controllably depends on the driving frequency and strength. Two LDSs either merge into one, or are both
annihilated. In both cases, we temporally resolve the collision dynamics and clearly observe the dissipative nature
of the interaction through analysis of the energy balance.

To better illustrate our experimental findings, we start our discussion by presenting numerical results. In
dimensionless form, the AC-driven damped NLSE reads
\begin{equation}
  \label{LLNAC}
  i\Psi_t+|\Psi|^2\Psi + \Psi_{xx} = -i\Psi + iSe^{i\Delta t}.
\end{equation}
This equation represents in our case the mean-field behavior of a Kerr resonator~\cite{lugiato_spatial_1987,
*haelterman_dissipative_1992, *wabnitz_suppression_1993}, but is also the small amplitude limit of the AC-driven
sine-Gordon equation~\cite{barashenkov_existence_1996, *barashenkov_existence_1999}. It has applications in
non-equilibrium systems ranging from plasma physics~\cite{kim_development_1974} to Josephson junctions
\cite{ustinov_solitons_1998}, highlighting the general applicability of our study. The equation can be cast into an
autonomous form by substituting $\Psi(x,t) = \psi(x,t) e^{i\Delta t}$,
\begin{equation}
  \label{LLN}
  i\psi_t+|\psi|^2\psi + \psi_{xx} = -i\psi + \Delta\psi + iS
\end{equation}
which will be used throughout this Letter.

Depending on the driving strength $S$ and its frequency $\Delta$, Eq.~\eqref{LLN} exhibits a range of solutions,
which have been extensively investigated~\cite{lugiato_spatial_1987, barashenkov_existence_1996,
*barashenkov_existence_1999}. Briefly, the simplest steady-state solutions are homogeneous (${\psi_{x}=0}$), and they
satisfy the well-known cubic steady-state equation $X = Y^3-2\Delta Y+(\Delta^2+1)Y$ with $X = |S|^2$ and $Y =
|\psi|^2$. The steady-state curve ($Y$ vs. $X$) is single-valued for $\Delta <\sqrt{3}$, whereas for $\Delta
>\sqrt{3}$ it assumes an S-shaped hysteresis cycle with three possible states. The latter range is of more relevance
to our experimental configuration~\cite{leo_temporal_2010}, and thus the focus of this Letter. Only two of the three
states that exist for $\Delta>\sqrt{3}$ are homogeneously stable (bistability): the negative slope branch is always
unstable. For $Y>1$, the upper branch exhibits a Turing-pattern instability (also known as modulation instability)
with respect to inhomogeneous perturbations, which can lead to the formation of a stationary periodic
pattern~\cite{lugiato_spatial_1987}. LDSs can manifest themselves under conditions of coexistence of a patterned
solution and a stable homogeneous solution (viz. $Y<1$). They can be understood to coincide with the patterned
solution over a finite region in $x$, and with the homogeneous solution elsewhere~\cite{lugiato_introduction_2003}.
\begin{figure}[t]
  \includegraphics[width=1\columnwidth]{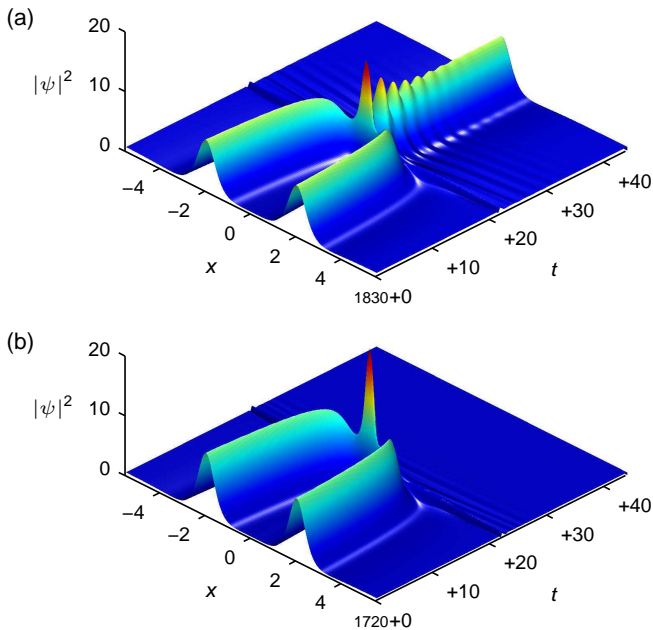}
  \caption{(color online). Numerically simulated dynamics of induced LDS interactions for $\phi(0) = 0.5~\mathrm{rad}$ and
    $x_0 = 30$. In (a) $[\Delta, S_0] = [2.91, 1.87]$ and two LDSs merge into one;
    in (b) $[\Delta, S_0] = [3.64, 2.10]$ and two LDSs annihilate one another.}
  \label{Simulation}
\end{figure}

We are interested in the dynamics that take place when two LDSs collide. Unlike conservative solitons,
widely-separated LDSs of the AC-driven NLSE are phase-locked to the driver, and thus all of them possess identical
traits (for given $X$ and~$\Delta$), including frequency ($\Delta$) and velocity. Accordingly, unassisted collisions
occur only when two LDSs are sufficiently close to interact attractively~\cite{cai_bound_1994,
*brambilla_interaction_1996}, yet such interactions are difficult to explore controllably. Inducing collisions by
suitably modulating the phase of the driver~\cite{rozanov_new_1992, *rosanov_diffractive_1993,
mcintyre_all-optical_2010} addresses that issue. Specifically, given $S(x) = S_0\exp[i\phi(x)]$, an LDS at
$x_\mathrm{L}$ will move towards the local maximum of $\phi(x)$ with a drift velocity of
$\mathrm{d}x_\mathrm{L}/\mathrm{d}t = \phi'(x_\mathrm{L})$~\cite{firth_optical_1996-1, jang_temporal_2014}. A
collision is thus observed when exciting, for example, two LDSs on opposite sides of a local maximum of
$\phi(x)$~\cite{mcintyre_all-optical_2010}.

To illustrate such induced collisions, we numerically integrate Eq.~\eqref{LLN} using the split-step Fourier method.
We assume a Gaussian driver phase profile $\phi(x) = \phi(0)\exp(-x^2/x_0^2)$. To create LDSs symmetrically
distributed about the phase maximum at $x = 0$, we use the initial condition $\psi(x,0) =
\sqrt{2\Delta}\,\bigl[\mathrm{sech}[\sqrt{\Delta}(x-x_\mathrm{L})]+\mathrm{sech}[\sqrt{\Delta}(x+x_\mathrm{L})]\bigr]$.
The initial LDS separation $2x_\mathrm{L} = 70$ is chosen to be much larger than their characteristic width ($\sim
1/\Delta$)~\cite{wabnitz_suppression_1993, barashenkov_existence_1996, coen_universal_2013} so as to avoid any
interactions during the transients leading to the LDS formation. Figures~\ref{Simulation}(a) and (b) show typical
results for two different sets of driver frequency $\Delta$ and strength $S_0$, as listed in the caption, and with
$\phi(0) = 0.5$~rad and $x_0 = 30$. These parameters are chosen to replicate our experiments. For clarity, the
figures neglect the initial portion of the simulation (which lasts for more than $t = 1700$), during which the two
LDSs slowly approach each other from their initial separation of~$2x_\mathrm{L}$. In both cases, it can be seen that
the LDSs drift towards each other until they are close enough to interact. The outcome of the collision is, however,
markedly different. Indeed, for $[\Delta, S_0] = [2.91, 1.87]$ the two LDS merge into one [Fig.~\ref{Simulation}(a)],
whilst for $[\Delta, S_0] = [3.64, 2.10]$ the intracavity field after the interaction is globally reduced to the
homogeneous solution, i.e., the two LDSs annihilate one another [Fig.~\ref{Simulation}(b)].

\begin{figure}[b]
  \includegraphics[width=1\columnwidth]{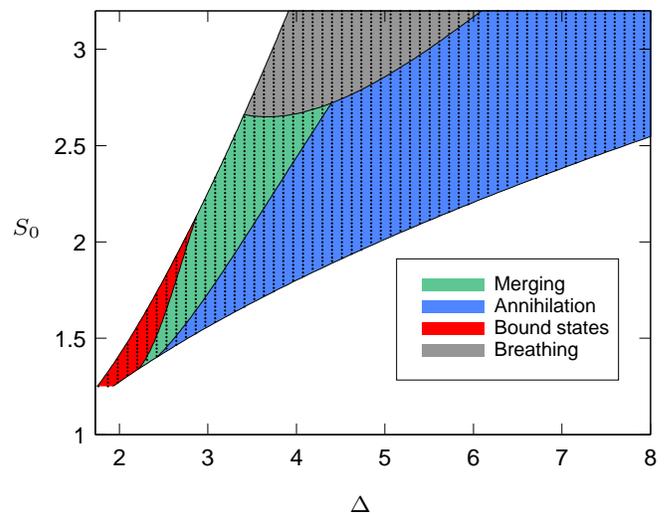}
  \caption{(color online). Results from numerical simulations illustrating the outcome of LDS interactions as a function of driving
    frequency $\Delta$ and strength $S_0$ for $\phi(0) = 0.5~\mathrm{rad}$, and $x_0 = 30$. Each solid dot
    represents a distinct simulation. No LDSs exist in the white-colorcoded area.}
  \label{phase_space}
\end{figure}

It is apparent that the interactions depend on the parameters of the driver. This has been numerically explored
further by systematically varying our four control parameters, $\Delta$, $S_0$, $\phi(0)$, and $x_0$, over a wide
range. We have found that $\Delta$ and $S_0$ mainly govern the outcome of the collision, while the phase modulation
parameters $\phi(0)$ and $x_0$ mostly determine the speed at which the LDSs approach each other, i.e., set the timing
of the collision. In Fig.~\ref{phase_space}, we summarize the observed outcome of the interaction as a function of
$S_0$ and $\Delta$ for the same driver phase modulation $\phi(x)$ as above. As can be seen, merging (green) and
annihilation (blue) occurs in clearly distinct, but adjacent, regions. For a given driving strength $S_0$, the system
favors annihilation over merging at higher driving frequencies. This can be related to the closer proximity to the
folding point at $\Delta_\mathrm{c} \sim \pi^2 |S_0|^2/8$, beyond which LDSs cease to exist in this
system~\cite{barashenkov_existence_1996}. Interestingly, in the area marked ``bound states'' no collision occurs.
Instead, the two LDSs form a stable bound state~\cite{barashenkov_bifurcation_1998, cai_bound_1994}: repulsive
interactions of the LDSs resist the drift induced by the driver phase modulation. Not surprisingly, this region
slightly grows at the expense of the ``merging'' region when a shallower phase modulation is used (the
merging/annihilation boundary is mostly unaffected). In the grey region, labeled ``breathing'', the individual LDSs
exhibit breathing as a result of an underlying Hopf bifurcation~\cite{barashenkov_existence_1999}. Their interaction
can lead either to merging or annihilation, depending on the phase of their breathing at the onset of the collision.
The bound-state and breather regimes will not be further discussed in this Letter because experimental limitations
currently prevent us from observing them.

We now describe our experimental configuration, implemented in the optical domain. Specifically, we induce
controllable LDS interactions in a coherently-driven passive optical fiber resonator that exhibits instantaneous Kerr
nonlinearity. In the high-finesse limit, this system is known to be governed by Eq.~\eqref{LLN}, with $\psi$
representing the slowly-varying envelope of the electric field~\cite{haelterman_dissipative_1992}. The LDSs of such
Kerr resonator have been observed experimentally before and are usually referred to as temporal cavity
solitons~\cite{leo_temporal_2010}. These are pulses of light that continuously circulate in the resonator, yet remain
stationary in a reference frame that is moving at the group velocity of the driving light in the fiber. The
transverse coordinate~$x$ in Eq.~\eqref{LLN} is thus a ``fast-time'' $x \rightarrow \tau$ that is defined in such a
reference frame and that allows to describe the temporal profile of the field envelope. In contrast, $t$ is a
``slow-time'' that describes changes in the field envelope over consecutive roundtrips around the resonator. The
normalization is such that dimensional time-scales $\tau'$ and $t'$ (units of~s) and the electric field envelope
$E(t', \tau')$  (units of~$\mathrm{W}^{1/2}$) are related to the dimensionless variables in Eq.~\eqref{LLN}
by~\cite{leo_temporal_2010}
\begin{align}
  t &= \alpha \frac{t'}{t_\mathrm{R}},
  & \tau &= \tau' \sqrt{\frac{2\alpha}{|\beta_2|L}},
  & \psi &= E\sqrt{\frac{\gamma L}{\alpha}}.
\end{align}
Here $t_\mathrm{R}$ is the roundtrip-time of the resonator, $\alpha$ is equal to half the percentage of total power
loss per round-trip, $L$ is the resonator length, and $\beta_2$ ($<0$) and $\gamma$ are, respectively, the anomalous
group-velocity dispersion and Kerr nonlinearity coefficients of the fiber. The driving strength~$S_0$ is related to
the power $P_\mathrm{in}$ of the continuous-wave (cw) laser driving the resonator as $S_0 = (P_\mathrm{in}\gamma L
\theta/\alpha^3)^{1/2}$, where $\theta$ is the intensity transmission coefficient of the coupler used to inject the
field into the resonator. Finally, $\Delta$ characterizes the frequency detuning of the cw driving laser at $\omega$
from the closest resonator resonance at~$\omega_0$, $\Delta \simeq t_\mathrm{R} (\omega_0-\omega)/\alpha$.

\begin{figure}[t]
  \includegraphics[width=1\columnwidth]{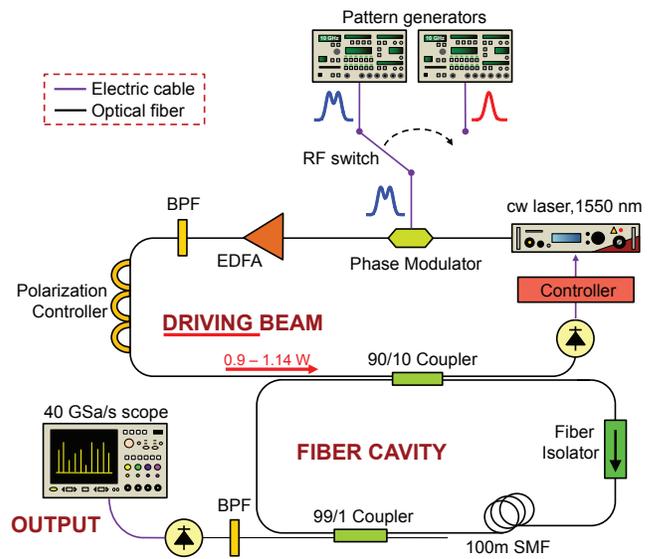}
  \caption{(color online). Experimental setup. cw: continuous-wave, EDFA: erbium-doped fiber amplifier, BPF: band-pass-filter,
    SMF: single-mode fiber.}
  \label{setup}
\end{figure}

A detailed schematic of our experimental setup is shown in Fig.~\ref{setup}. Overall, it is similar to the one used
in~ \cite{jang_temporal_2014}. As a coherent driver, we use a narrow linewidth cw laser at 1550~nm wavelength, which
is amplified up to $1.14$~W using an erbium-doped fiber amplifier (EDFA) before being coupled into the resonator by a
90/10 fiber coupler ($\theta = 0.1$). Noise accumulated during the amplification stage is mostly removed with an
optical band-pass filter (BPF). The resonator is composed of 100~m of standard silica single-mode fiber (SMF), with
$\beta_2 = -21.4~\mathrm{ps^2/km}$ and $\gamma = 1.2~\mathrm{W^{-1}km^{-1}}$. It also incorporates an optical
isolator to prevent resonance of stimulated Brillouin scattering radiation, and a 99/1 fiber coupler through which
the intracavity dynamics are monitored with a fast photodiode and a real-time oscilloscope. The overall finesse of
the resonator was measured to be $\mathcal{F} = \pi/\alpha \sim 21.5$. The BPF at the 1\,\% output filters out the
homogeneous cw background that coexists with the LDSs, thereby improving the signal-to-noise ratio of our
data~\cite{leo_temporal_2010}. The resonance frequencies of our optical fiber ring generally exhibit fluctuations due
to environmental perturbations. To maintain a fixed $\Delta$, we therefore actively actuate the driving laser
frequency to follow any changes in the resonances, by locking to a set level the optical power reflected off the
resonator input~\cite{leo_temporal_2010, jang_ultraweak_2013, jang_temporal_2014}. Changing the lock point allows us
to controllably adjust~$\Delta$, but we remark that the accuracy with which we can do so is insufficient to explore
the formation of bound states since they manifest themselves over a narrow range of driver frequencies (see
Fig.~\ref{phase_space}). In this context, we also note that, with our current configuration, we are unable to reach
power levels required to explore interactions of breathing LDSs.

To controllably induce LDS interactions, we phase modulate the resonator driving field with a 10~GHz electro-optic
modulator. The modulator is driven by one of two 10~GHz programmable pattern generators, selected with an electronic
switch. The pattern generators are synchronized to each other by a single external clock, such that the repetition
rate of their output patterns is identical to the resonator free-spectral range. The first generator (left in
Fig.~\ref{setup}) is configured to produce a pattern of two 130~ps full-width-at-half-maximum electronic pulses with
200~ps separation. These are fed to the phase modulator in the initial stage of the experiment. During that stage, we
mechanically perturb the resonator, which results in the direct excitation of two LDSs at the two phase maxima
\cite{jang_writing_2015}. After the LDSs are stably formed and trapped at the maxima \cite{jang_temporal_2014}, we
activate the electronic switch and the phase modulator feed is abruptly changed (within a few nanoseconds) to the
output of the second pattern generator. That generator is set to produce a pattern made up of a single pulse whose
delay is adjusted to lie halfway between the two pulses generated by the first generator. Accordingly, the LDSs in
the resonator find themselves approximately symmetrically positioned about the new single maximum of the phase
profile. As in the simulations of Fig.~\ref{Simulation}, the LDSs thus start drifting towards that maximum,
interacting once sufficiently close to each other. Note that the new driver phase profile takes a few photon
lifetimes ($\sim 1\ \mu$s) to get imprinted inside the resonator after the switch, but that transient is negligible
in regards of the interaction time.

\begin{figure}[t]
  \includegraphics[width=1\columnwidth]{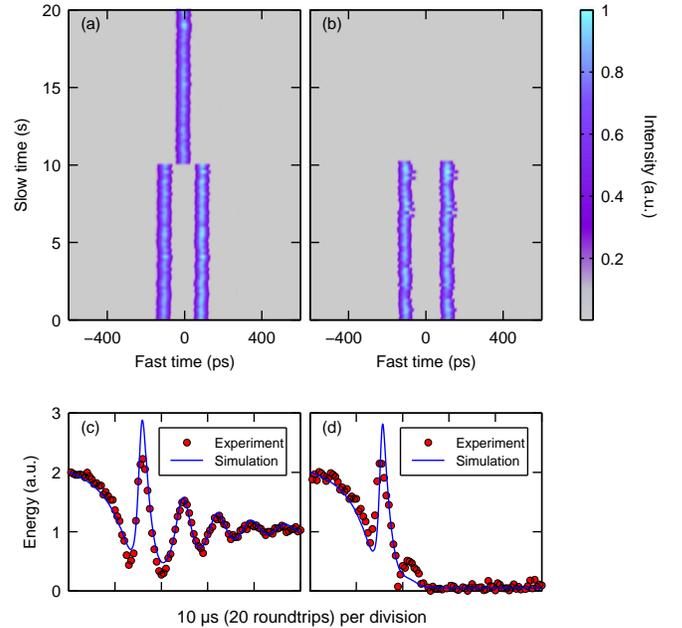}
  \caption{(color online). (a, b) Experimental density plots showing the evolution of the intracavity temporal intensity profile as
    two LDSs (a) merge into one, and (b) annihilate each other. The successive traces are recorded with a real-time
    oscilloscope at 1~frame/s. (c, d) The roundtrip-to-roundtrip evolution of the total intracavity energy during LDS
    (c) merging and (d) annihilation. The experimental data is represented by red circles, and results from numerical
    simulations are shown as blue solid lines. }
  \label{colormap}
\end{figure}

The temporal intensity profile of the intracavity light measured at the 1\,\% output of the resonator is recorded by
the oscilloscope (triggered by the pattern generators) as we operate the switch. Successive recordings are vertically
concatenated and shown as density plots in Figs.~\ref{colormap}(a) and (b), which have been obtained for $[\Delta,
S_0] = [2.91, 1.87]$ and $[3.64, 2.10]$, respectively. The first 10~s of the measurements are very similar: two LDSs
with 200~ps separation are stably trapped at the maxima of the phase pulses defined by the first pattern generator.
After switching to the single phase pulse pattern (which occurs at $t \simeq 10$~s) and inducing the LDS interaction,
a single LDS is seen to remain for $[\Delta, S_0] = [2.91, 1.87]$ whilst both disappear when $[\Delta, S_0] = [3.64,
2.10]$. These results are indicative of merging and annihilation, respectively, which is in agreement with numerical
simulations. Indeed, the simulation results in Fig.~\ref{Simulation} use the very same parameters as the experiments
here. Yet, these results are limited by the slow 1~frame/s acquisition rate of the oscilloscope which does not reveal
the transient energy balance dynamics. We have thus also recorded the roundtrip-to-roundtrip evolution of the
intracavity energy on the real-time oscilloscope. Experimental results for merging and annihilation are shown as red
circles in Figs.~\ref{colormap}(c) and (d), respectively, superimposed with results from numerical simulations (blue
solid lines). Here we only show the portion of the dynamics corresponding to the final stages of the interaction (the
energy stays approximately constant during the slow approach of the two LDSs) and we have normalized the energy such
that a single isolated LDS carries the energy 1~a.u. We have also post-processed the numerical simulation results,
extracted from data shown in Fig.~\ref{Simulation}, to take into account the BPF at the resonator output as well as
the limited bandwidth of our photodetector. The results in Fig.~\ref{colormap}(c,~d) clearly confirm that merging and
annihilation occur in our experiment and more importantly the dissipative nature of the interactions. For $[\Delta,
S_0] = [2.91, 1.87]$ the energy falls from two to one, implying merging; for $[\Delta, S_0] = [3.64, 2.10]$ the
energy falls from two to zero, implying annihilation.

These results represent, to the best of our knowledge, the first realization of controllable interactions of
localized dissipative structures. Our study also provides the first quantitative analysis of such interactions in an
AC-driven nonlinear Schr\"odinger system, and more generally, in any nonlinear dissipative system near the 1\,:\,1
resonance. We have numerically identified a diversity of interaction scenarios for different parameters of the system
driver. Experiments performed in an optical resonator show unequivocal evidence of possible selection of LDS
interaction by the operator from merging to annihilation.

We acknowledge financial support from the Marsden fund of the Royal Society of New Zealand. M. Erkintalo also
acknowledges support from the Finnish cultural foundation.

\end{document}